\PassOptionsToPackage{table,svgnames}{xcolor}
\documentclass[sigconf]{acmart}

\AtBeginDocument{%
  \providecommand\BibTeX{{%
    \normalfont B\kern-0.5em{\scshape i\kern-0.25em b}\kern-0.8em\TeX}}}

\acmConference[arXiv]{}{June 14th}{2024}

\usepackage{algorithm}
\usepackage{tabularx} 

\begin{document}

\title{Carbon-Aware End-to-End Data Movement}

\vspace{15mm}
\author{Jacob Goldverg, Hasibul Jamil, Elvis Rodriguez and Tevfik Kosar}
\affiliation{University at Buffalo, Buffalo, NY 14260\\
\{jacobgold, mdhasibu, elvisdav, tkosar\}@buffalo.edu
\country{}
}
\email{}

\vspace{15mm}

\begin{abstract}
The latest trends in the adoption of cloud, edge, and distributed computing, as well as a rise in applying AI/ML workloads, have created a need to measure, monitor, and reduce the carbon emissions of these compute-intensive workloads and the associated communication costs. The data movement over networks has considerable carbon emission that has been neglected due to the difficulty in measuring the carbon footprint of a given end-to-end network path. We present a novel network carbon footprint measuring mechanism and propose three ways in which users can optimize scheduling network-intensive tasks to enable carbon savings through shifting tasks in time, space, and overlay networks based on the geographic carbon intensity.
\end{abstract}
\vspace{3mm}


\maketitle

\vspace{3mm}

\section{Introduction}

Data has become the foundation of today's IT-dependent world, and the plethora of data generated by AI workloads, scientific applications, social media, and e-commerce fuels large-scale data analytics systems. As a result, data transfer over the Internet has been increasing each year exponentially and has already exceeded the zettabyte scale~\cite{Cisco_2019}.
With the increased data generation rate, the data movement's carbon footprint is becoming an overwhelmingly critical problem, especially for HPC and Cloud data centers.
It is estimated that information and communication technologies will use between 8\% - 21\% of the world's electricity by 2030 ~\cite{belkhir2018assessing}.
The share of communication networks in the total IT power consumption is around 43\%~\cite{challengesinNetworks}. 
Data transfers over the Internet consume more than a hundred terawatt-hours of energy, which costs 20 Billion US dollars annually. This trend has motivated considerable work to reduce the energy consumption of networking at all layers, including the hardware, software, protocol, and applications.

Despite the advances in networking technologies, sending data over networks is still very costly in terms of energy consumption and carbon footprint. The researchers found that sending hard drives between collaborating institutions would be many orders of magnitude less carbon-emitting than transferring the data over communication networks~\cite{aujoux2021estimating}. 
And, due to increased energy costs pushed by constantly increasing traffic volumes, current network energy costs of telecommunication service operators in developing countries already span between 40\% and 50\% of provider operational expenditures~\cite{lorincz2019greener}.

There has been a considerable amount of work focusing on power management and energy efficiency in hardware and software systems~\cite{Brooks:2000:WFA:339647.339657, rawson2004mempower, zedlewski2003modeling, gurumurthi2002using, contreras2005power, economou2006full, fan2007power, rivoire2008comparison, koller2010wattapp, hasebe2010power, vrbsky2013decreasing} and on power-aware networking~\cite{Katz_2008, Mahadevan_2009, Greenberg_2009, Heller_2010, Goma_2011, alan2014energy, alan2015energy, guner2018energy, di2019cross,  rodolph2021energy, jamil2022energy, greenNFV}. More recently. a consensus among major IT companies and federal agencies has shifted the focus to the environmental impact of this vast power consumption and underscored the urgency to decrease carbon emissions during the entire lifecycle of computing, including networking and data movement~\cite{NSF-WSCS-2024}.

\begin{figure}[h]
    \centering
    \includegraphics[width=\linewidth]{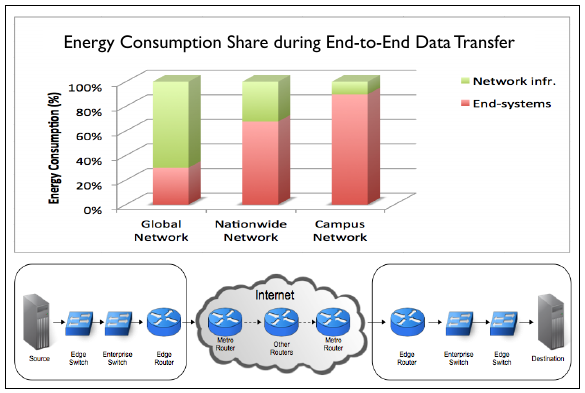}
    \caption{Energy consumption share of the end systems vs network infrastructure during end-to-end data transfer. This ratio depends on the number of network devices (i.e., routers, switches, etc.) between the end systems, and how much power each device consumes.}
    \label{fig:endsystems}
\end{figure}

Despite the broad range of research in power management techniques for the networking infrastructure, there has been little work focusing on reducing the carbon emissions of end-to-end data movement, including the emissions at the end systems (i.e., sender and receiver nodes) during active data transfers. A study shows that approximately 25\% of total electricity consumption during the end-to-end data transfers occur at the end-systems on a global (intercontinental) network, whereas this number goes up to 60\% on a nationwide network and up to 90\% on a local area
network~\cite{alan2015energy}. This ratio depends on the number of network devices (i.e., routers, switches, hubs, etc.) between the sender
and receiver nodes and how much power each device consumes, as shown in Figure ~\ref{fig:endsystems}.

\vspace{4mm}

Measuring, monitoring, and optimizing end-to-end data movement for carbon efficiency is a challenging task. Although there has been some recent work on the carbon efficiency of the networking infrastructure during data movement~\cite{carbonAwareNetworking, carbonAwareRouting, tabaeiaghdaei2023carbon}, the carbon emissions at the end systems (i.e., sender and receiver nodes) during data movement have been long ignored. This makes it difficult to understand the end-to-end carbon intensity of data movement.

This paper makes several novel contributions to the field: (1) it presents a method to measure the carbon footprint of end-to-end data movement on a given network path by computing the carbon intensity for each hop on the path as well as for the end systems; (2) it discusses the potential of scheduling data movement in time and space to reduce the carbon footprint of data transfers; and (3) it shows the need for fine-grained host monitoring and combining this with carbon reporting of network paths.

\section{Background and Related Work}

The internet stack is collectively comprised of many components starting from end systems that is either the sender or receiver, into network switches and routers and finally the destination end system. From the application level users are completely abstracted away from the network infrastructure and exposed minimal amounts of information via the operating system. TCP based protocols are client driven thereby users on the client side are able to obtain networking metrics such as: latency, RTT, dropped packets and throughput. Furthermore Linux tools such as Traceroute enables users to measure the network path used for communication if the destination server and ISP's do not drop or mask ICMP packets \cite{traceroute}. Unfortunately routers and network switches do not report utilization metrics to users in the application level making it near impossible for users to understand buffer sizes, and ports used along the network path, hence users are left with very trivial networking metrics in the application layer.

Due to the large amount of components in internet technology, measuring the carbon intensity is non-trivial and requires specific parts of the internet stack to be measured independently when constructing an end-to-end carbon intensity. Carbon intensity is interpreted as the amount of carbon dioxide emissions produced per unit of computing power or data processed. Networks can be broken down into hardware and software components. On the hardware side, there are network switches and routers that collectively do not expose any form of metrics related to carbon intensity. Metrics related to resource utilization are exposed and then referencing the models data sheet we are able to get an estimate on energy consumption where we can then correlate the energy draw to the carbon-intensity. The software side of a network is much more challenging to measure as we generally require hardware support to measure electrical consumption.

Carbon-aware routing is a recently explored topic that would make Border Gateway Protocol (BGP) route internet traffic based on the carbon intensity of network regions. By disabling and routing traffic through different regions instead of just QoS and other metrics ISPs commonly consider, large carbon savings can be achieved for general internet users \cite{carbonAwareRouting}. Furthermore, due to the challenges in observing the power consumption of routers and networking components, a traffic engineering approach becomes highly feasible \cite{carbonAwareRouting}.

The scheduling of compute-bound tasks is a topic heavily studied in varying domains. In big data centers, jobs are executed based on user priority such that the priority determines if the job can be shifted in time and space as well as to the extent. Shifting jobs in time to maximize: green energy utilization, throughput, and electrical load utilization of data centers is being done today \cite{carbonExplorer}. 

Carbon-aware networking faces many challenges due to the lack of kernel and network metrics reported in the application space. The sum cost of all networking components that comprise the network stack is essential when determining the carbon footprint imposed when an end-systems network utilization varies \cite{carbonAwareNetworking, challengesinNetworks}. Unfortunately, prior work in making networks carbon-aware has ignored the carbon emission during data transfer at the end systems. 

In mobile cloud environments, job executions may be paused and transferred to another device if there is insufficient battery life to complete the entire job \cite{offloadingInMobileEdgeComputing}. This process, known as compute offloading, requires an overlay network where a central node (or the network) manages available nodes that can take over the job, allowing it to continue on a different device. Due to energy constraints on mobile devices, checkpointing a job and then running it on a device with sufficient power resources enables the task to be completed without wasting previously computed work\cite{offloadingInMobileEdgeComputing}.

In modern 5G networks, there is a significant focus on offloading compute-intensive tasks to nearby cloud providers. This practice aims to improve both battery life and throughput of tasks that are limited by computational resources \cite{pbe_cc, offloadingInMobileEdgeComputing}. Green NFV (Network Function Virtualization) investigates the optimal placement and scheduling of virtual network functions (VNFs) to maximize throughput, energy efficiency, and resource utilization \cite{greenNFV}. Additionally, there has been extensive research on optimizing big data transfers to enhance throughput and end-system utilization \cite{onedatashare, AppLevelOpt, blaze, kosar2005data, kosar2019energy, jamil2022reinforcement, jamil2022energy, bufferTuningEsma}.

There are three common concepts introduced in the covered past work that enable the ability for data transfer offerings to optimize for carbon intensity. Shifting a scheduled file transfer in time allows the file transfer node to have a different carbon intensity along the same network path by starting the job when renewable resources are more readily available. The file transfer node running the transfer can also be changed during the job execution, thereby using a different network path which results in a greener transfer, we refer to as using an overlay network. Finally, due to the duplicity of files that are distributed geographically, making it possible for file transfer nodes (FTNs) to use a file source that has a greener network path, we refer to this as shifting in space.

In this work, we emphasize the ability for data movement tools to optimize for carbon intensity, instead of just throughput and energy efficiency, through carbon aware scheduling. 
Existing efforts have been made in developing approaches to making the internet greener but have failed in including the cost of the end-system \cite{carbonAwareNetworking, challengesinNetworks}. 
We emphasise the importance of measuring the carbon-intensity of end systems as well as the network path utilized for the file transfer.

\section{Carbon Intensity of Data Path}

With the centralization of data as well as the growth in data size, the demand for fast and efficient file transfers has become essential for domain scientists to conduct research. Due to this, various efforts have been made in optimizing file transfer performance either through utilizing multiple network paths \cite{blaze}, tuning TCP parameters \cite{blaze, AppLevelOpt, onedatashare, globus} as well as scheduling the file transfer when bandwidth is more available \cite{storkcloud}. With such efforts being made in throughput and end-to-end file transfer efficiency, we would like to present how file transfer platforms can optimize for carbon intensity. Furthermore, we would like to demonstrate the importance of measuring the carbon intensity from the file source to the file destination.

Offering end-to-end carbon-aware file transfers is dependent on: 
\begin{enumerate}
    \item End system monitoring and reporting of fine-grained network and host metrics;
    \item Measuring of IP addresses from file source to destination;
    \item The ability to measure the longitude and latitude position of each IP address and look up the carbon intensity per location;
    \item A scheduler capable of ingesting such information and then launching transfers when carbon savings.
\end{enumerate}

\begin{table}[htb]
    \centering
    \caption{End-system metrics collected for monitoring}
    \label{tab:metrics}
    \rowcolors{3}{gray!25}{white}
    \begin{tabularx}{\columnwidth}{X|X|X}
        \hline
        \textbf{Host Metrics} & \textbf{Network Metrics} & \textbf{Transfer Metrics} \\
        \hline
        Core Count & Drop Out & Job Uuid\\
        Free Memory & Drop In & Source Latency \\
        Max Memory & Error In & Job Size\\
        Memory & Error Out& {Transfer Node ID} \\
        {\scriptsize Min}CPU Frequency & Dst Latency& Buffer Size\\
        {\scriptsize Max}CPU Frequency & Src Rtt& Parallelism\\
        {\scriptsize Current}CPU Freq& Dst Rtt& Concurrency \\
        CPU Architecture & NIC MTU & Pipelining \\
        CPU Utilization & Network Interface&Bytes Received\\
         & Packet Sent & Bytes Sent \\
         & Packet Received &  \\
         & NIC Speed &  \\
         & Read Thrpt  &  \\
         & Write Thrpt  &  \\
        \hline
    \end{tabularx}
\end{table}

\subsection{End-system Monitoring}
Various libraries and tools exist to expose kernel, CPU, and network metrics into the application space of an end system. Monitoring CPU and system power consumption is a hardware-dependent task that relies on specialized hardware. Mobile systems today typically provide a file to monitor the power draw of different components in the device. Android provides an Energy Profiler that captures the power consumption of the antenna, CPU, and GPS. Unfortunately, most consumer hardware does not support fine-grained power monitoring \cite{androidPower}. Users have some options provided to them when using Linux. Perf is a common utility that provides power consumption of some processors, and Intel provides RAPL to give users direct joules consumed per core \cite{intelRapl, linuxPerf}. To capture the networking metrics, users can use specific libraries such as Psutil, netstat, and parsing the Linux files related to TCP, UDP, and raw socket information \cite{netstat, psutil}. A non-exhaustive list of crucial metrics a tool such as this must report is listed in Table \ref{tab:metrics}.

\subsection{Discovering a Network Path}
Typical Linux-based solutions use either ICMP or UDP to discover the network path from sender to receiver. This is done by manipulating the Time to Live (TTL) of a packet triggering a response from a router with its IP address. This process is done until the destination IP address has been reached. Packet manipulation libraries such as Scapy can be used to do this manually, or using a tool like Traceroute is sufficient to discover the network path \cite{traceroute, scapy}. Users have numerous options to discover the network path all depending on the operating system, programming language, and available libraries.

\subsection{Geolocating and Measuring the Carbon Intensity of a Network Path}
To find the longitude and latitude coordinates for a list of public IP addresses, geolocation databases are used. These databases can be either offline or online, and there are many free options available. One such option is IP-API, which provides an online geolocation database with free API access. This database is regularly updated to ensure the most recent mapping of IP addresses to geographical positions \cite{ipApi}. Once the coordinates are obtained, services like Electricity Maps or WattTime can be used to retrieve live or historical carbon intensity information for the corresponding geographic regions \cite{electricityMaps, wattTime}.

\begin{figure}[t]
  \centering
  \includegraphics[width=\linewidth]{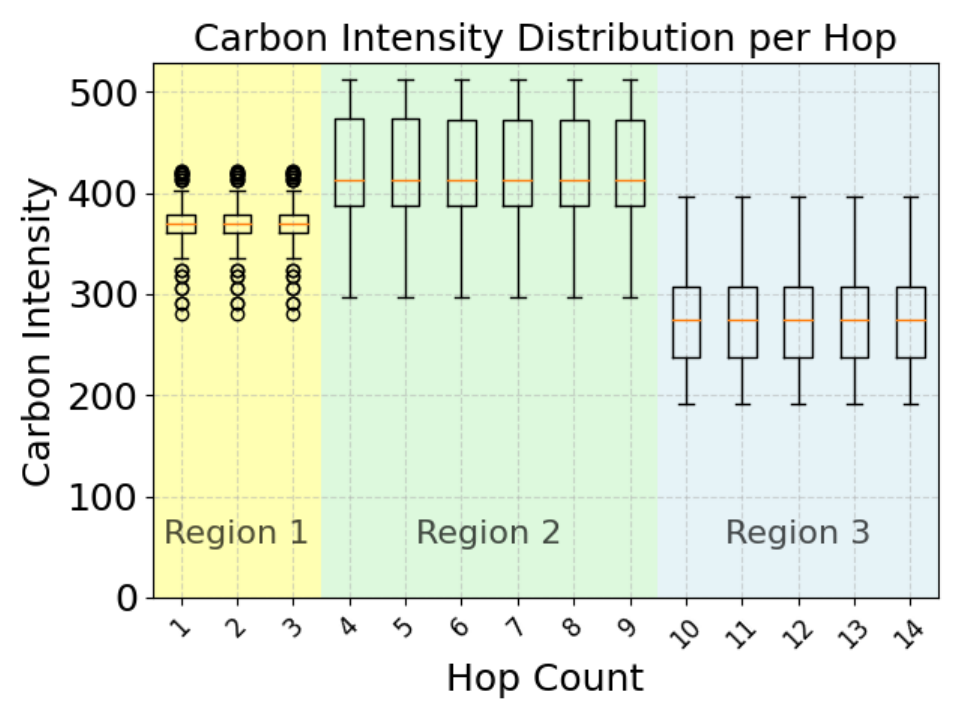}
  \caption{Carbon Intensity of IP addresses from UC to TACC}
  \label{fig:traceRouteCITaccUC}
\end{figure}

We created a boxplot to visualize the process as shown in Figure \ref{fig:traceRouteCITaccUC}. This plot covers a 51-hour period and reveals natural clusters and variations of carbon intensity among IP addresses that share the same electrical regions. We notice that the hops are all grouped into natural regions through the carbon intensity values, making the network path groupable based on the energy provider.

\subsection{Carbon Intensity of a File Transfer}
Throughput and carbon intensity are both changing values over time, hence it is crucial to track both numbers over the duration of the entire file transfer. With this, we propose a trivial formula to calculate the carbon intensity per byte moved over a network. 
\begin{equation}
  \text{$carbon score$} = \frac{\text{$bytes$}}{\text{$CI$} \times \text{$duration$}}
\end{equation}
where $bytes$ is the total number of bytes transferred, $CI$ is the average carbon intensity of the file transfer, and $duration$ is  the total amount of time the file transfer took in seconds. Hence, we interpret this metric as the carbon intensity per bit per second. This formulation allows us to understand the performance as well as the imposing carbon cost of the file transfer.

By measuring the carbon intensity throughout a file transfer, tools can calculate the total average carbon intensity per bytes per second. After the transfer finishes, the job can be evaluated based on both the achieved throughput and the carbon impact of the file transfer. Using this simple heuristic and bandwidth prediction, it becomes possible to estimate the carbon impact a file transfer will have at a certain time. This prediction would take into account the previously viewed throughput of jobs given the same file source and destination as well as the application parameters for the transfer. Thereby it will allow a scheduler to estimate the carbon impact and throughput beforehand.

Due to there being no current open-source tool that offers the discussed features, we chose to build our own. Pmeter \cite{pmeter} is an open-source tool that: uses the ICMP protocol to discover the network path, IP-API \cite{ipApi} to to geolocate each IP address, Electricity Maps \cite{electricityMaps} to discover the carbon intensity per IP, and provides such this feature set agnostic of the underlying operating system monitoring tool. 
\section{Experimentation}
Chameleon Cloud (CC)\cite{chameleonCloud} is an HPC cloud like provider offering baremetal hardware for research purposes. This is preferred for our initial experimentation as it allows us to avoid the cost of virtualization and shared hardware resources, as well get operating system metrics imposed by strictly the utilized applications. We also use an Apple M1 Macbook Pro that is hosted in a DIDC lab. Two nodes are deployed in CC that are geographically separated, in TACC we deploy an Nginx server with the firewall fully disabled, in UC, and on the macbook we deploy the OneDataShare FTN with Pmeter installed on the system \cite{onedatashare, pmeter}. We depict the utilized hardware for our experimentation in Table \ref{tab:nodes_hardware_spec}.

\begin{table*}
  \caption{Nodes used for Carbon Intensity Measurement}
  \label{tab:nodes_hardware_spec}
  \rowcolors{1}{gray!25}{white}
  \begin{tabular}{c|c|c|c|c|c|c}
   \toprule
    {\bf Provider} & {\bf Site} & {\bf Flavor} & {\bf CPU} & {\bf RAM} & {\bf NIC} & {\bf OS} \\
    \midrule
    Chameleon Cloud & TACC & Baremetal & Cascade Lake & 192 GiB & 10 Gbps & Ubuntu 22 \\
    Chameleon Cloud & UC & Baremetal & SkyLake & 192 GiB & 10 Gbps & Ubuntu 22  \\
    DIDCLab & University at Buffalo & Baremetal & Apple M1 & 14.9 GiB & 1.2 Gbps & mac OS Sonoma \\
      \bottomrule
\end{tabular}
\end{table*}

\subsection{Shifting in Time}

\begin{figure}[t]
    \centering
    \includegraphics[width=\linewidth]{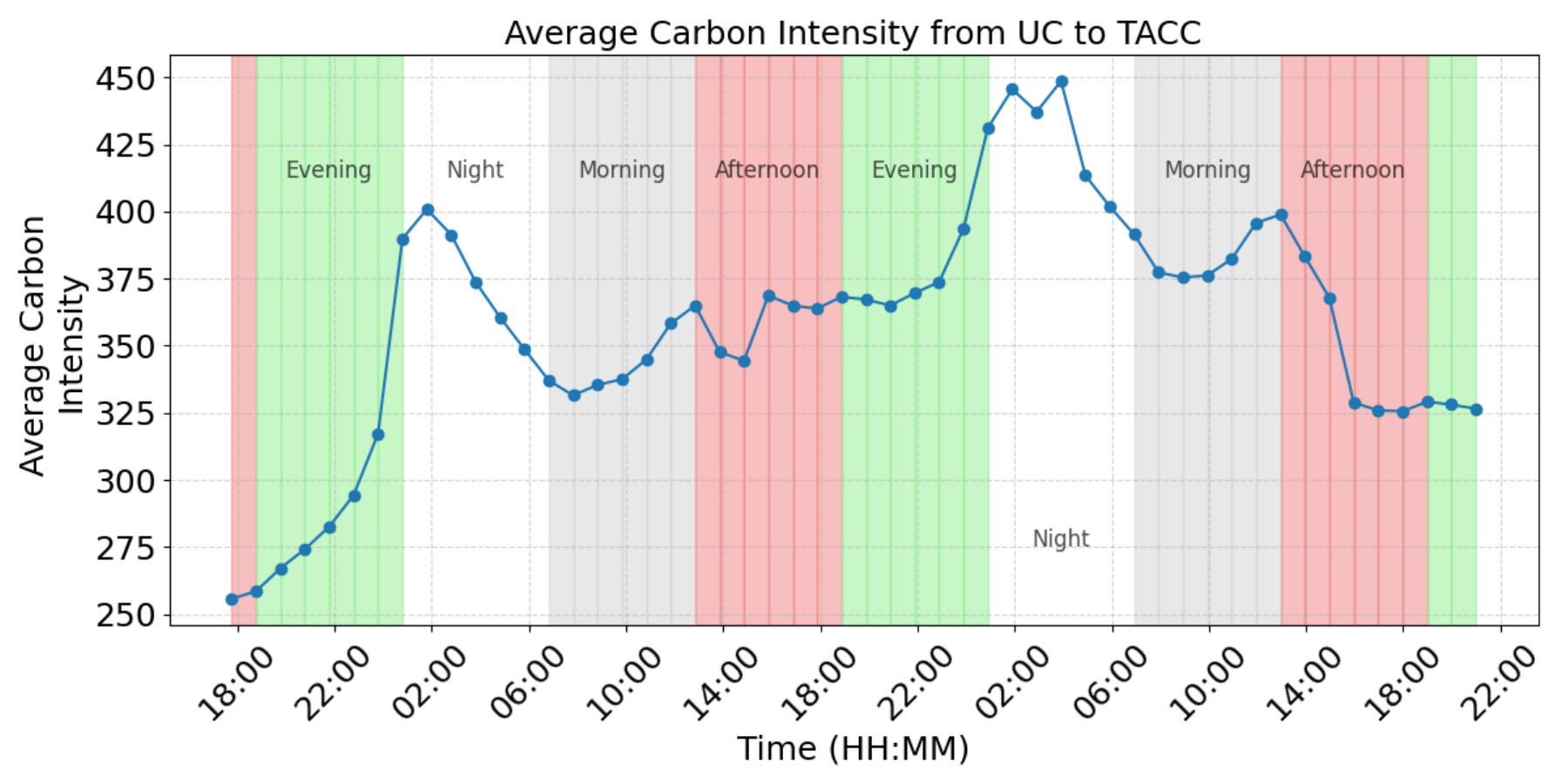}
    \caption{Average Carbon Intensity at 1-Hour Intervals from UC to TACC, April 14-16, 2024}
    \label{fig:thirtyIntervalCI}
\end{figure}

Shifting a file transfer in time utilizes the same file source, destination, and FTN but simply changes the time at which the file transfer begins. To motivate the impact of shifting a file transfer in time, we would like to visually show the average carbon intensity from UC to TACC. Every hour, Pmeter is run and we visualize the average carbon intensity output between the two nodes in Figure \ref{fig:thirtyIntervalCI}. 
We observe that the minimum average carbon intensity over the $51$ hour period is $255.714$ where the maximum is $488.6$, implying by simply shifting a job in time a carbon aware scheduler can achieve nearly \~2x in carbon savings.

\subsection{Shifting in Space}

\begin{figure}[t]
    \centering
    \includegraphics[width=\linewidth]{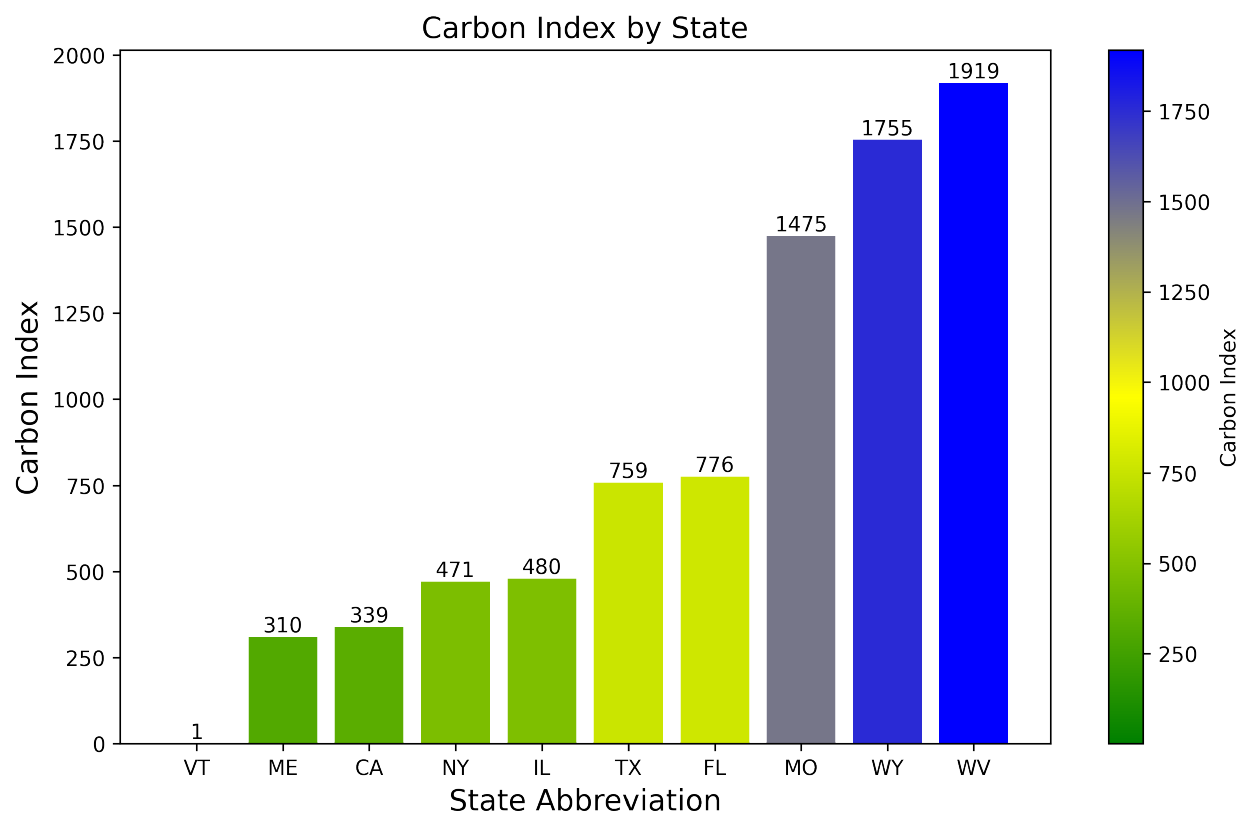}
    \caption{Carbon Index of 10 states in the US}
    \label{fig:carbon_idx_states}
\end{figure}

Shifting in space as previously described is when there is data duplicated on many servers, similar to Content Delivery Networks (CDNs). A carbon-aware scheduler picks a server as the file source with the lowest associated carbon intensity to conduct the file transfer. A further motivation of this is to visualize the associated carbon index of various states throughout the United States \cite{emissionIndex}, we visualize the carbon index of ten states in Figure \ref{fig:carbon_idx_states}.

Determining the appropriate source file server is crucial in minimizing the overall carbon intensity as states in the US have varying carbon index's. In the most extreme case we see that Wyoming has a carbon index of $1919$ where Vermont has a score of $1$. This implies that if the same dataset was duplicated in both states, then a scheduler that chooses Vermont as the file source would lead to $1919x$ in carbon savings strictly due to selecting the dataset residing in a greener location.

\subsection{Using an Overlay Network of File Transfer Nodes}

\begin{figure}[t]
    \centering
    \includegraphics[width=\linewidth]{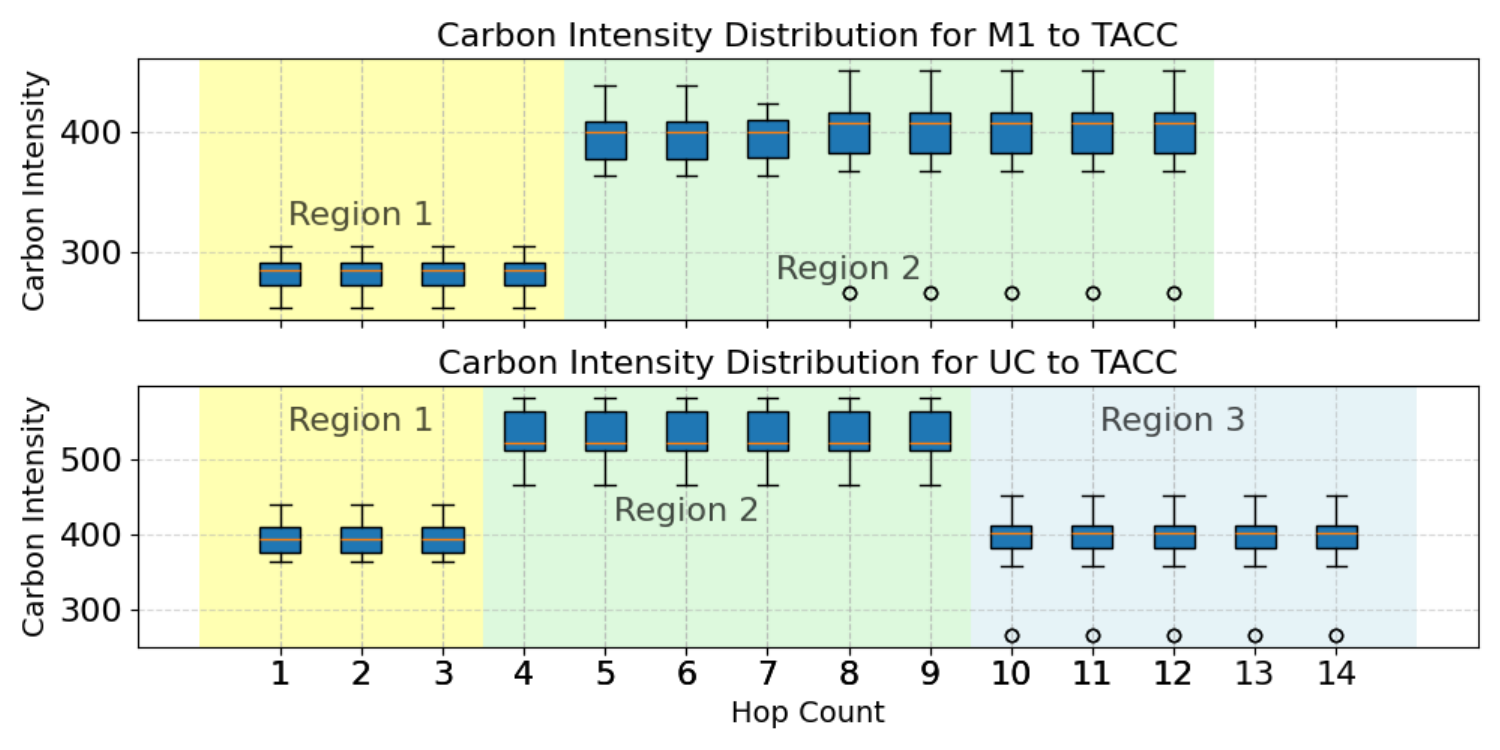}
    \caption{Measuring the Carbon Intensity from UC and M1 to TACC}
    \label{fig:overlay_network_box_motivation}
\end{figure}

Overlay networks provide the capability to select the FTN responsible for processing a file transfer job and allow for the migration of work among FTNs when there are resource limitations. For carbon-aware file transfer scheduling using overlay networks is possible as choosing an FTN with the lowest carbon intensity to execute the file transfer. Moreover, if a predefined carbon intensity threshold is exceeded, the remaining work can be migrated to an FTN with a lower carbon intensity. To illustrate the impact of overlay networks on carbon-aware file transfers, we compare the average carbon intensity of the network path where the file source is hosted at TACC, while both UC and a MacBook host the FTNs that would download a dataset. In Figure \ref{fig:overlay_network_box_motivation}, we can see that the MacBook in the DIDC lab is better positioned to process the file transfer job due to the lower carbon intensity of the entire network path and its path to TACC having fewer hops.

\section{Future Work and Conclusion}

The emphasis on file transfer performance and efficiency has opened up potential directions for making file transfers more carbon-efficient. This study introduced three techniques for scheduling file transfers to reduce carbon costs. Future work should integrate these techniques with established metrics, such as throughput, latency, and fair-share utilization, alongside the new carbon intensity metrics. Given the complexity of incorporating a highly stochastic metric like carbon intensity, it is crucial to balance this with past work.

In end-to-end data transfer tuning\cite{TCP_Pipeline, twoPhaseDynamicThroughput, kosar2019energy, bufferTuningEsma, esmaGridftp, onedatashare} and end-to-end data pipeline optimization~\cite{deelman2006makes, bahsi2007conditional, kougka2018many}—topics that have been extensively researched should also consider carbon intensity. An immediate area of research could explore the relationship between throughput and carbon intensity, striving to balance them effectively during file transfers. Further investigation is also needed on how application parameters influence the carbon impact of a file transfer. Using power models that estimate the energy consumption of end systems based on memory, CPU, and network utilization in conjunction with the carbon intensity of the end system will be essential for calculating total carbon costs\cite{alan2015energy}.

Moreover, with Service Level Agreements (SLAs), platforms can offer users the choice to set preferred carbon footprints and performance levels for their jobs, thus promoting green computing initiatives. This approach ensures high user satisfaction by balancing QoS with carbon efficiency, thereby offloading the responsibility of carbon savings of a job to the user and not just the platform.

Cloud providers, with their geographically distributed data centers, stand to benefit significantly from the presented techniques with the goal of carbon savings. Meta has already implemented and utilized shifting jobs in time to maximize the utilization of green energy \cite{carbonExplorer}. By using overlay networks, placing jobs in data center sites that are physically closer to the file source and have a greener network path to the required dataset, data centers can reduce carbon intensity associated with network I/O. Testbeds such as Fabric and Chameleon Cloud can be utilized to study how positioning of jobs at certain sites can potentially increase throughput and reduce carbon intensity by bringing the compute task closer to the data \cite{chameleonCloud, fabric-2019}. 

CDNs and caches are widely used to enhance QoS and reduce latency for end users. CDNs are regularly updated to provide the latest content during off-peak hours, thereby reducing server downtime. However, there has been limited research on the carbon intensity of CDNs and their communication costs due to these updates. This area of study could benefit from time and space shifting as well as using overlay networks. Time shifting could involve updating the CDNs when carbon intensity is lowest, while space shifting could optimize the file source location of the newest content for minimal carbon impact. Overlay networks may help determine the update sequence of CDNs to further reduce the carbon intensity of the network path.

In conclusion, as we delve deeper into the carbon implications of networked systems and file transfers, it becomes imperative to integrate carbon cost considerations into future computing research. This study has shown that application-level tools and libraries can play a great role in monitoring and revealing the increasing carbon costs associated with file transfers over network paths.

\section*{Acknowledgements}
This project is in part sponsored by the National Science Foundation (NSF) under award number OAC-2313061.

\bibliographystyle{ACM-Reference-Format}
\bibliography{references}

\end{document}